# Large-Scale Geospatial Processing on Multi-Core and Many-Core Processors: Evaluations on CPUs, GPUs and MICs

Jianting Zhang
Department of Computer Science
The City College of New York
New York, NY, USA
jzhang@cs.ccny.cuny.edu

Simin You
Dept. of Computer Science
CUNY Graduate Center
New York, NY, USA
syou@gc.cuny.edu

*Abstract—* Geospatial Processing, such as queries based on point-to-polyline shortest distance and point-in-polygon test, are fundamental to many scientific and engineering applications, including post-processing large-scale environmental and climate model outputs and analyzing traffic and travel patterns from massive GPS collections in transportation engineering and urban studies. Commodity parallel hardware, such as multi-core CPUs, many-core GPUs and Intel MIC accelerators, provide enormous computing power which can potentially achieve significant speedups on existing geospatial processing and open the opportunities for new applications. However, the realizable potential for geospatial processing on these new hardware devices is largely unknown due to the complexity in porting serial algorithms to diverse parallel hardware platforms. In this study, we aim at experimenting our data-parallel designs and implementations of point-to-polyline shortest distance computation (P2P) and point-in-polygon topological test (PIP) on different commodity hardware using real large-scale geospatial data, comparing their performance and discussing important factors that may significantly affect the performance. Our experiments have shown that, while GPUs can be several times faster than multi-core CPUs without utilizing the increasingly available SIMD computing power on Vector Processing Units (VPUs) that come with multi-core CPUs and MICs, multi-core CPUs and MICs can be several times faster than GPUs when VPUs are utilized. By adopting a Domain Specific Language (DSL) approach to exploiting the VPU computing power in geospatial processing, we are free from programming SIMD intrinsic functions directly which makes the new approach more effective, portable and scalable. Our designs, implementations and experiments can serve as case studies for parallel geospatial computing on modern commodity parallel hardware.

I. INTRODUCTION

Geo-referenced or geospatial data are universal in many science and engineering disciplines, ranging from environmental sciences to intelligent transportation systems and location dependent services. The increasingly popular GPS devices and GPS-enabled smartphones, and the advances in environmental sensing and modeling technologies, have generated huge amount of geospatial data. For example, more than 2.7 billion GPS points have been collected and made available to the public by global contributors to Openstreemap [1]. Thousands of GPS-equipped buses in New York City (NYC) are sending a GPS location every 30 seconds or so [2] which results in millions of points a day. The Global Biodiversity Facility (GBIF) data portal hosts 400+ million species occurrence records [3] which contain rich distribution patterns of millions of documented species and are crucial for global biodiversity analysis. These point locations are meaningful when they are aligned to global, regional and urban infrastructures, such as administrative zones, ecological regions and street networks.

While the performance of traditional disk-resident spatial database systems based on serial algorithms are far from satisfactory in processing large-scale geospatial data [4], the newly emerging parallel hardware provides considerable computing power for speeding up geospatial processing on large-scale data. Our recent work on developing data parallel techniques for spatial data management on GPUs have demonstrated significant potentials [4,5], but the performance comparisons reported in these studies are limited to multi-core CPUs on the same machine that hosts the GPU for the experiments. The data parallel designs make it easy to port our GPU code to new generation of multi-core CPUs equipped with VPUs and the newly available Intel MIC devices [6]. In addition, both multi-core CPUs and GPUs are increasing their number of cores and cache sizes, and, CPUs also have wider SIMD width (from 128-bit SSE to 256-bit AVX) [7]. While our previous experiments have demonstrated significant performance gains of GPUs over previous generation of CPUs [4, 5], it is interesting to perform more comprehensive comparisons on new generation multi-core CPUs, GPUs and MICs.

Our work is also motivated by the observation that, while both Intel MICs are increasingly used in many scientific and engineering disciplines, most of existing applications focus on numeric computation which typically involve matrix manipulations that have regular memory access patterns (e.g., [8,9]). In contrast, many spatial operations require significant irregular memory accesses. For example, the numbers of vertices in polygons and polylines may vary significantly and there is significant branching in both point-to-polygon shortest distance computation (denoted as P2P) and point-in-polygon topological test (denoted as PIP). We believe our experiments on Intel MICs, in addition to GPUs and multi-core CPUs, are valuable in understanding the relatively pros and cons of using these commodity parallel hardware for scientific and engineering applications. Our technical contributions can be summarized as follows:

- We have successfully ported GPU-based designs and implementations of P2P and PIP spatial operations to Intel MIC accelerators based on Intel Thread Building Blocks [10] and the Intel SPMD Program Compiler (ISPC [11]) to utilize multi-core CPUs and their VPUs, respectively.
- We have performed extensive experiments on seven commodity parallel hardware platforms, including three multi-core CPUs, three GPUs and a MIC accelerator.
- We report our experiment results and provide in-depth analysis on how parallel hardware architectures and configurations may significantly affect performance of application programs, and discuss how a DSL approach can be effective in utilizing SIMD computing power on modern parallel processors.

The rest of the paper is arranged as the following. Section II introduces background, motivation and related work. Section III presents the designs and implementations of the two spatial operations after introducing the spatial filtering pre-processing step. Section IV provides experiment settings and results with discussions. Finally Section V is the conclusion and future work.

II. BACKGROUND, MOTIVATION AND RELATED WORK

While the combination of architectural and organizational enhancements led to 17 years of sustained growth in performance at an annual rate of 50% from 1986 to 2003, the growth rate has dropped to 22% per year from 2003 to 2010 due to the combined power, memory and instruction-level parallelism problem [7]. The hardware changes have significant impacts on software and applications. In geospatial processing, while the majority of commercial and open source software are still based on serial algorithms on uniprocessors and disk-resident systems, there are growing research and applications that target at exploiting parallel processing power of multi-core CPUs and many-core accelerators, including GPUs and the Intel MICs [6]. However, the performance of geospatial operations on the new commodity parallel hardware is still unclear, especially for operations on vector spatial data where data accesses are largely irregular and there are significant divergences in control logics, when compared with raster-based operations where efficient parallel designs and implementations are available based on dense matrix manipulations. In this study, we aim at filling the gap by experimenting P2P and PIP spatial operators to help understand the achievable level of performance on commodity hardware.

It is beyond our scope to provide a comprehensive review of available operations on geospatial data but we refer to the Open Geospatial Consortium (OGC) Simple Feature Specification (SFS) [12] for the list of defined operations that are more query-oriented and a Geographical Information System (GIS) textbook [13] for operations that focus more on geospatial analysis and interactive visualization. Query-oriented spatial operations are typically put into a Spatial Join framework [14]. It is well known that spatial joins have two phases, i.e., filtering and refinement. The filtering phase relies on various spatial indexing structures to filter out a large portion of candidate pairs to be joined while the refinement phase computes spatial relationships among filtered candidate pairs. The P2P and PIP spatial operations can be considered as two special cases of computing spatial relationships. In this study, we will be focusing on the refinement phase. Given a pair of a set of points and their neighboring polylines or polygons that are derived in the spatial filtering phase, for each point, we want to find its nearest polyline within range $R$ (for P2P), or, the polygon it falls within (for PIP). There are several parallelisms that can be mapped to different parallel hardware (e.g., multi-core CPUs, GPUs and MICs) at different levels (e.g., multi-processor, thread-blocks, SIMD elements) and the design and implementation details will be presented in Section III.

In addition to exploring parallelisms for spatial processing on large-scale geospatial data, we are also interested in understanding how different generations of commodity hardware with similar architectures affect end-to-end performance for our applications. Ideally parallel designs can exploit the inherent parallelisms in applications that can automatically scale across hardware generations. However, the achievable speedups are often significantly limited by many other factors, such as caches, VPU SIMD widths, memory bandwidths, efficiencies of runtime libraries and granularities of parallelisms. Towards this end, in this study, we have experimented our implementations on three types of multi-core CPUs and three types of GPUs with different hardware configurations for the two spatial operations using real large-scale data. As detailed in Section IV, our experiments do suggest that, the performance of parallel implementations improve across different generations of parallel hardware in general, although fine-tuning for specific hardware may be needed for higher performance. In a way similar to relying on CPU speed improvements for better performance automatically in the "serial age" back to 1980s and 1990s, we expect that parallel designs can also benefit from parallel improvements of modern hardware in a somewhat automatic manner (to a certain degree) in the "parallel age". This might warrant investments of parallelizing traditional serial algorithms whose initial cost may be high with respect to re-design and re-implementation.

The Intel MIC accelerators (e.g., Xeon Phi 3120A used in this study) are especially interesting as they have features of both multi-core CPUs and many-core GPUs. MICs are similar to multi-core CPUs due to their origination and the majority of CPU code can be relatively easily ported to MICs without major changes. However, MICs are also similar to GPUs as they have much more processing cores, much larger memory bandwidths, larger SIMD widths, smaller caches than multi-core CPUs, and, currently support in-order execution only. As detailed in Section IV, for some experiments, MICs have achieved comparable or even better performance when compared with GPUs. However, for some other experiments, MICs are inferior to multi-core CPUs. We have also observed that the performance of single-core on MICs can be up to 10-15 times lower than a single-core on multi-core CPUs when VPUs are not used on the same machine, despite that the clock frequency difference is less than 2.5X (2.6 GHZ vs. 1.1GHZ) in our experiments. This brings an interesting question on what features of multi-core CPUs should be kept on MICs and what should not for better end-to-end performance on MICs with comparable hardware budgets. We hope the experiments of our domain-specific applications can be useful in this regard.

Despite VPUs have been part of CPUs for a long time and instruction sets (e.g., MMX, SSE and AVX [7]) have been provided for SIMD-based computation, to the best of our knowledge, there are no previous implementations of spatial query processing using VPUs, possibly due to the complexity and non-portability of using assembly-alike SIMD intrinsic functions. As VPU SIMD widths are getting larger and larger (e.g., 8-way for Ivy Bridge multi-core CPUs and 16-way for Xeon Phi MICs), it becomes more beneficial to exploit VPU SIMD computing power in terms of cost-effectiveness. The ISPC compiler [11] has provided a DSL approach which allows users write scalar-alike code, in a way similar to CUDA or OpenCL code for GPUs, to utilize VPUs more easily. The compiler will translate the scalar code to parallel code by automating looping through 1D arrays and calling appropriate SIMD intrinsic functions. Although spatial operations that involve irregular data accesses and have complex control logics are unlikely to fully utilize VPU SIMD computing power, the compiler makes it possible to write portable parallel code across different generations of VPUs in a cost-effective way. We note that, while VPUs on MICs have comparable SIMD width (512-bit, 16-way) with that of GPUs (1024-bit, 32-way), unlike a GPU thread that has its own (but slow) registers and can tolerate control divergence to a certain degree in a graceful manner, all SIMD elements on VPUs share the same registers and do not support control divergence natively. Control divergences may require much more complex steps (e.g., using mask registers) which may significantly reduce parallelisms and hence performance. While it certainly takes more research to decide whether incorporating some GPU features on VPUs will be beneficial in a wide range of applications, we consider our experiments can serve as domain case studies on designing future VPUs, in addition to help understand the achievable performance improvements when deciding whether to exploit SIMD computing power on VPUs. Previous works on comparing the performance of multi-core CPUs, GPUs and MICs for different application domains (e.g., [15, 16]) either did not utilize VPUs [15] or relied on compiler-based auto-vectorization [16]. In contrast, we have adopted a DSL approach to utilizing VPUs which is more convenient and effective in incorporating complex semantics in expressing parallelisms.

### III. PARALLEL DEISGNS AND IMPLEMENTATIONS OF TWO SPATIAL OPERATIONS

#### A. Data-Parallel Spatial Query Processing Framework

As mentioned earlier, a large number of spatial indexing techniques have been proposed over the past few decades. Although most of them are serial algorithms for historical reasons, more and more parallel designs and implementations on commodity parallel hardware are becoming available. In our previous works, we have explored GPU-based R-Trees [17], Quad-Trees [5] and Simple Flat Grid Files (SFG) [4] for spatial indexing and spatial filtering. In this study, we will use SFG for spatial filtering to pair up sets of points with their neighboring polylines or polygons for refinements. As shown in the top-left part of Fig. 1, by sorting points based on their cell identifiers, we can derive all the points that fall within the cells. As detailed in [4], this can be realized by using a few simple parallel primitives that are supported by major parallel libraries. The grid cell identifier vector (VGC) will be accompanied by the VGI vector that indexes the points that fall within the grid cells whose identifiers are stored in VGC. The vector of polygon/polyline identifiers (VPC) and the vector of polygon/polyline vertex indices (VGI) serve the same purposes for polygons or polylines. For each of polylines/polygons, we can derive its Minimum Bounding Boxes (MBBs) and map it to the same grid as points, as shown in the top-right part of Fig. 1. Clearly, a MBB will intersect with one or more grid cells and this is the simple geometric foundation for grid-based spatial indexing. As such, vector VPP is used to maintain the one–to-many relationship between polygons or polylines and grid cells.

The purpose of spatial filtering is to pair points with neighboring polylines or polygons. To support locating the nearest polyline for a point (x,y) within a query window width of *R*, as shown at the middle-left part of Fig. 1, it is sufficient to examine all the polylines whose expanded MBBs intersect with the point. Assuming the MBB of a polyline is (x1,y1,x2,y2), for a group of points in a grid cell, the necessary condition for at least one of the points that are at most *R* distance away from a polyline is that the grid cell that the point falls within intersects with the expanded MBB of the polyline (x1-R,y1-R,x2+R,y2+R) [4].

For PIP, a simple observation is that, a grid cell may be inside, intersect and outside of a polygon (polygons are allowed to have holes as shown in the top-left part of Fig. 1). No points in a grid cell that is outside of a polygon can be within the polygon and thus can be safely removed from the test. If a cell is completely within a polygon, then all points will be in the polygon without requiring further test. As such, only points within the grid cells that intersect with polygon boundaries need point-in-polygon tests, which is likely to dominate the whole computing process. We will be focusing on these points in our experiments in Section IV.

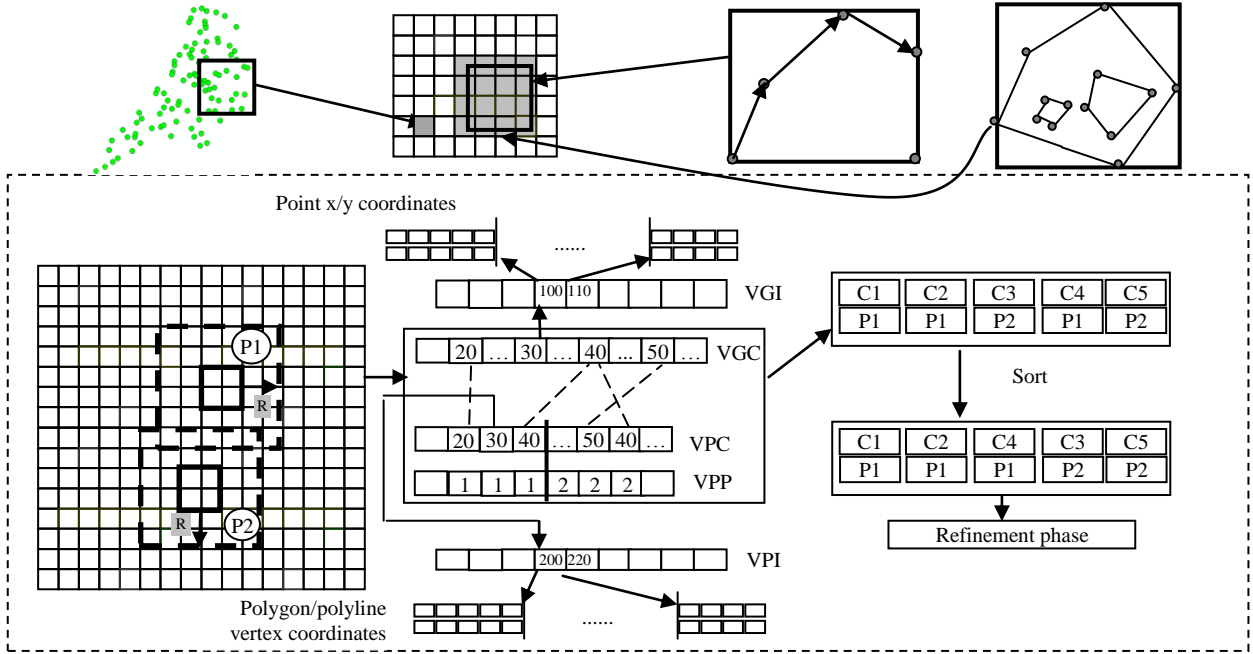

Fig. 1 Framework of Spatial Query Processing using Data Parallel Designs and Parallel Primitives-Based Implementations

In both cases, the expanded MBBs of polylines and the MBBs of polygons need to be rasterized based on the same grid tessellation for points for spatial filtering purposes. The key part of spatial filtering is binary search based on common grid cell identifiers (middle part of Fig. 1) and. We refer to [4] again for design and implementation details on both MBB rasterization and binary-search based cell identifier pairing. After sorting on the binary search results, each grid cell will be paired up with a set of MBB identifiers. Using the cell and MBB identifiers, it is straightforward to retrieve the x/y coordinates of points in the cell and polyline/polygon vertices for both P2P and PIP. The details will be provided in the next two subsections (Section III.B and III.C).

It is clear that using a smaller grid cell size will result in larger number of (point cell, MBB) pairs. This subsequently reduces false positives which in turn lowers the computation overheads at the spatial refinement phase at the cost of increasing the computation costs in the filtering phase, in addition to using more memory for storage (e.g., all the five vectors listed in the middle of Fig. 1 are likely to be larger). This is a well-known tradeoff in spatial query processing. Since our focus in this study is to evaluate the performance on different hardware instead of indexing geospatial data optimally, we choose an appropriate grid cell size based on our previous studies without further optimization.

As the designs on spatial indexing and spatial filtering largely involve element-wise operations on 1D vectors which are suitable for parallel primitives–based implementations, we consider our framework for spatial query processing data-parallel and fine-grained, in comparison with task-parallel approaches that rely on tightly coupled global structures (e.g., priority queues) and the parallelisms that are being explored are typically coarse.

*B. P2P Design and Implementation*

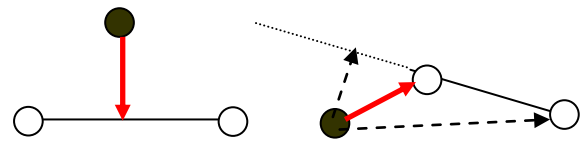

*Fig. 2* Illustration of Two Cases of Shortest Distance between a Point and a Line Segment

By definition, a polyline has multiple line segments and the distance between a point and a polyline, as illustrated in Fig. 2, is canonically defined as the shortest distance between the point and all the line segments in a polyline. When the point is projected to a line segment, if the projection point falls between the two ends of the line segment, the point-to-line-segment distance is the distance between the point and the projection point (perpendicular distance); otherwise the point-to-line-segment distance will be the shorter of the distances between the point and the two ends of the line segment [4].

While the serial implementation of computing the shortest distance between a single point and a single polyline is straightforward, it takes some thoughts on data layout and data parallelization schemas for point datasets with multiple point grid cells and polyline datasets also with multiple polylines. Please note that the number of points in point grid cells and the number of vertices in polylines may vary significantly. Different from traditional Object-based representation on uni-processor CPUs that abstracts away these variations by representing points and polylines polygons as objects, as shown at the top and bottom part of Fig. 1, we use auxiliary vectors to store numbers of points and vertices explicitly. By performing a parallel prefix sum (scan), the starting position of points in a cell, or vertices in a polyline, can be computed either dynamically or through a pre-processing step. As such, at the finest parallelization level, we can assign a pair of (point, vertex) to a SIMD unit to compute their distance and all SIMD units can work in parallel, as each unit knows where to load point/vertex coordinates by adding the starting position and SIMD unit specific offset.

Since the spatial filtering phase returns pairs of (CID,{PID}) where CID represents a grid cell identifier and PID represents a polyline identifier that is at most $R$ distance away from at least one of the points in CID, the parallelisms among (CID,{PID}) pairs can be considered as the top level parallelism. In our implementation, we assign a (CID,{PID}) pair to a thread block on GPUs and a batch of $K$ pairs as a TBB task on multi-core CPUs and MICs [10]. We further assign a point in the grid cell CID to a thread on GPUs. Similarly, we assign a point to a SIMD element when VPUs are used on multi-core CPUs and MICs. If VPUs are not available, a CPU thread just simply processes all the points in its outmost loop. In both cases, all points loop through vertices of one or multiple polylines represented by {PID}. While using SIMD intrinsic functions directly would require an explicit loop when the number of points in the grid cell is larger than the VPU SIMD width (currently 4-8 on multi-core CPUs and 16 on MICs), ISPC allows arbitrary long virtual SIMD length expressed as a *foreach* loop, in a way similar to CUDA where the virtual SIMD length is set to thread block size.

While we refer to [4] for more details on the design and implementation of the P2P operation on both multi-core CPUs and GPUs, here we would like to compare the design with an alternative one which is assigning a polyline vertex to a SIMD unit (a GPU thread or a VPU SIMD element) and let the SIMD units loop through all points. Clearly, the PIP control logic needs to be implemented in a single looping step, i.e., computing geometric distances, identifying the two cases illustrated in Fig. 2, and calculating the minimum distance. While this is doable using CUDA as cross SIMD element operations are well supported through parallel libraries, it is quite difficult on VPUs. This is because only very limited support on cross SIMD element operations (e.g., min/max) within physical SIMD elements are currently available and supporting generic cross SIMD element operations in software are non-trivial on VPUs. As such, we consider our original design a more viable solution based on the current support from both existing parallel libraries and hardware. In addition, the design also makes the implementation on GPUs and VPUs very similar. This makes it easy to understand and maintain the code, and, opens the possibility for future integration as well.

*C. PIP Design and Implementation*

Similar to P2P, the output of the spatial filtering phase is pairs of (CID,{PID}) for PIP where CID represent grid cell identifier and PID represent polygon identifier whose MBB intersect with the grid cell. Note that we do not need to expand polygon MBBs by distance $R$ in PIP. Again, we assign points to GPU threads and VPU SIMD elements and let SIMD units loop through vertices of one or more polygon rings in parallel. As detailed in [5], our implementation of point-in-polygon test on GPUs is based on the well known ray-crossing algorithm (Fig. 3) by adopting existing efficient serial implementation [18]. The GPU implementation is ported to ISPC code by using the same DSL approach outlined in Section III.B.

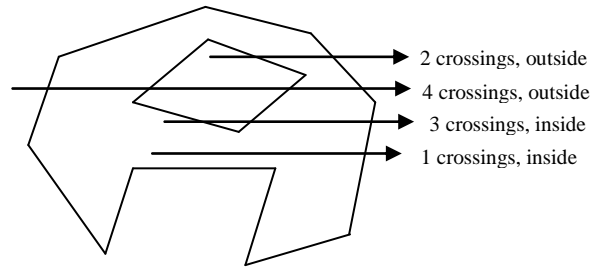

Fig. 3 Illustration of Ray-Crossing Algorithm for Point-In-Polygon Test (Polygon with One Hole)

IV. EXPERIMETNS AND EVALUATIONS

*A. Data, Hardware and Experiment Setups*

We use several real world datasets for our experiments. For P2P, we use the pickup locations of taxi trips in NYC in the first 6 months of 2009 as the point dataset and the NYC LION street network as the polyline datasets. Finding the shortest distance between taxi pickup locations and road segments with a certain distance has practical applications. If a taxi pickup location cannot be aligned to any road segments with a distance (say 100 feet), this pickup location is likely to be an outlier or the underlying road network is outdated. By computing the distributions of shortest distance between pickup locations and road segments, it is possible to understand GPS accuracies and their relationship with surrounding environments, assuming taxis can only pickup passengers along road segments. The number of street segments (polylines) is 147,011 and the number of total vertices is 352,111. The number of pickup locations is 84,035,490, i.e., nearly half a million a day.

For PIP, we use global ecological regions from World Wild Fund (WWF) as the polygon dataset which has 14,458 complex polygons and the total number of vertices is 4,045,460, i.e., about 280 vertices per polygon. Note that many of these polygons are complex with at least one hole which means the polygons have multiple rings. Furthermore, these complex polygons can also be non-convex. We use a GBIF global species occurrence dataset dump in 08/02/2012 with 375+ million species occurrences records. We extract the longitude and latitude coordinates from these records and use them as our point dataset for experiments. Since the full point dataset is too big for GPUs and MICs, in order to compare across CPUs, GPUs and MICs, we have extracted approximately 50 million points from species that have large numbers of occurrences in the dataset.

We have used three types of Intel Xeon CPUs, namely dual quad-core E5405 released in the fourth quarter of 2007 (Q4'07), dual quad-core E5520 released in the first quarter of 2009 (Q1'09), and dual 8-core E5-2650 V2 released in the first quarter of 2012 (Q1'12). The CPUs came as part of middle range workstations (~$5000) when the machines were purchased within a year or so after the CPUs were first released. The three Nvidia GPUs that we have used for experiments are Nvidia Quadro 6000 released in the third quarter of 2010 (Q3'10), Nvidia Tesla C2050 released in the third quarter of 2011 (Q3'11), and Nvidia GTX Titan released in the first quarter of 2013 (Q1'13). Note that Tesla C2050 and Quadro 6000 are based on Nvidia Fermi architecture while GTX Titan is based on the most recent Kepler architecture. Finally, the Intel Xeon Phi 3120A device is a low-end one of Intel's MIC-based architecture and is first released in second quarter of 2013 (Q2'13). More hardware features, including (micro)-architecture, number of cores, processor clock rate, SIMD width and memory bandwidth, are listed in table I. The purpose of stating release dates of these processors explicitly is to help understand how hardware with similar architecture improvements may "automatically" improve the performance of parallel designs and implementations as discussed in Section II.

Table 1 List of Major Specifications of 7 Parallel Processors for Experiments

|  | CPU1 | CPU2 | CPU3 | MIC | GPU1 | GPU2 | GPU3 |
|---|---|---|---|---|---|---|---|
| Model | Xeon E5405 | Xeon E5520 | Xeon E5-2650 V2 | Xeon Phi 3120A | Quadro 6000 | Tesla C2050 | GTX Titan |
| (Micro) Architecture | Core (Harpertown) | Nehalem (Gainestown) | Ivy Bridge (EP) | Knight Corner | Fermi | Fermi | Kepler |
| #of Cores | 2*4 | 2*4 | 2*8 | 57 | 14*32 | 14*32 | 14*192 |
| Clock Frequency | 2.00 GHZ | 2.26 GHZ | 2.60 GHZ | 1.10 GHZ | 1.15 GHZ | 1.15 GHZ | 0.88 GHZ |
| SIMD Width | 4 | 4 | 8 | 16 | 32 (warp) | 32 (warp) | 32 (warp) |
| Memory bandwidth | 2*10.8GB/s | 2*25.6 GB/s | 2*59.7 GB/s | 240 GB/s | 144GB/s | 144GB/s | 288.4GB/s |

The P2P and PIP designs are first implemented using CUDA 5.5 on GPUs. To port the implementations to multi-core CPUs and MICs, Intel TBB 4.2 [10] and Intel ISPC 1.6 [11] are used for thread-level and SIMD-level parallelization on CPUs and VPUs, respectively. For all compilations, we use O2 for optimization as we have found that using O3 may actually decrease performance in several cases. Since it is possible to choose between using single core or multiple cores and choose to use or not to use VPUs independently, there are four configurations on multi-core CPUs and MICs. We use SC to denote single core without SIMD, MC to denote multi-core without SIMD, SC+SIMD to denote single core with SIMD, and MC+SIMD to denote multi-core with SIMD, respectively. We put GPU results under MC+SIMD category as all multi-processors are used and SIMD is part of GPU computing.

*B. Results of P2P Shorest Distance Computation*

The P2P (point-to-polyline shortest distance computation) results are shown in Table 2 and the five speedups, i.e., MC over SC, MC+SIMD over SC+SIMD, SC+SIMD over SC, MC+SIMD over MC, and the overall speedup calculated as MC+SIMD over SC, are listed in Table 3. All the three CPUs have demonstrated excellent speedups which are nearly linear with the numbers of cores. The overall speedups are 29X for older generation CPUs (CPU1 and CPU2 with 8 cores and 4-way SSE SIMD) and 43X for newer generation CPUs (CPU3 with 16 cores and 8-way AVX SIMD). Comparing with SIMD speedups (SC+SIMD over SC and MC+SIMD over MC) for CPU1 and CPU2, which are about 3.7 (out of 4) and close to linear, these speedups are a little lower for CPU3 which are only 3.1-3.3 (out of 8). We suspect that the low SIMD speedups for CPU3 can be due to memory bandwidth contentions at CPU3's processing rates which are much higher than those of CPU1 and CPU2 based on their runtimes.

When comparing the runtimes among CPU1, CPU2 and CPU3, we are surprised to see that CPU3 performances much better than CPU1 and CPU2. Since CPU3 only has 2X cores and moderately higher clock rate, which are 1.3X over CPU1 and 1.15X over CPU2, we expected to see 2X-4X speedups, instead of 23X-44X from the measured runtimes. We have repeated the experiments several times on the machines with CPU1, CPU2 and CPU3 and have verified the correctness of the results and consistency over multiple runs. Furthermore, we also have performed the same set of experiments on another machine that is identical with the same machine with CPU1. All the additional experiments have provided similar results as we have reported in Table 2. While we are still in the process of fully understanding the excellent performance on CPU3 from an application perspective, we suspect that the higher memory bandwidth and architectural improvements of CPU3 (Ivy Bridge V2) may contribute to the excellent performance. When comparing with the results of PIP experiments to be reported in Section IV.C (next subsection) where CPU3 only achieves less than 5X speedups over CPU1/CPU2 (which are more expected), the computation in P2P experiments is much lighter and the performance is more sensitive to memory bandwidths as well as caches. The higher memory bandwidth (59.7 GB/s per socket) and larger L3 cache (20 MB per socket) on CPU3 might be among the contributing factors

Table 2 List of Runtimes for Point-To-Polyline Experiments for all Seven Processors (in Milliseconds)

| Configuration | CPU1 | CPU2 | CPU3 | MIC | GPU1 | GPU2 | GPU3 |
|---|---|---|---|---|---|---|---|
| SC | 209411.14 | 206011.55 | 8273.15 | 178292.23 | | | |
| SC+SIMD | 57465.05 | 55751.30 | 2497.05 | 11116.38 | | | |
| MC | 26399.84 | 25778.37 | 593.70 | 1528.14 | | | |
| MC+SIMD | 7198.85 | 6969.37 | 189.04 | 108.59 | 383.77 | 385.92 | 338.9 |

Table 3 List of Speedups for Point-To-Polyline Experiments for Three CPUs and MIC

| | Configuration | CPU1 | CPU2 | CPU3 | MIC |
|---|---|---|---|---|---|
| Multi-core Speedup | SC/MC | 7.93 | 7.99 | 13.93 | 116.67 |
| | (SC+SIMD)/(MC+SIMD) | 7.98 | 8.00 | 13.21 | 102.37 |
| SIMD Speedup | SC/(SC+SIMD) | 3.64 | 3.70 | 3.31 | 16.04 |
| | MC/(MC+SIMD) | 3.67 | 3.70 | 3.14 | 14.07 |
| Overall Speedup | SC/(MC+SIMD) | 29.09 | 29.56 | 43.76 | 1641.88 |

The performance of the MIC accelerator is also interesting as the runtimes are dramatically improved from single-core without SIMD to multi-core with SIMD. The speedups for MC over SC and MC+SIMD over SC+SIMD using all the 228 threads, as shown in Table 3, are 110X and 62X, respectively. Although the speedups are above the number of physical cores (57), we note that the MIC accelerator allows 4-way hardware threading which is effective to hide latencies of slow processors on MICs. Our experiments support the effectiveness of hardware threading in this case. It is also interesting to see that the speedups due to SIMD are 16X and 14X for SC+SIMD over SC and MC+SIMD over MC, respectively. Given that the SIMD width on the MIC is 16-way, the high speedups seem to be questionable at first place but can be explained as follows.

We have also repeated the experiments several times and verified the correctness of the results and confirmed the consistency of the performance among multiple runs, although the variations are larger on the MIC accelerator than those on multi-core CPUs despite only a single job is running on the MIC accelerator. Since ISPC did not generate object code directly on MICs (which is different for multi-core CPUs) and source code is first generated before using Intel ICC compiler for source code compilation, we are able to look into the generated source SIMD code for MICs. It seems that certain SIMD-specific optimizations are applied by the ISPC compiler which makes the ISPC code potentially much faster than CPU code that does not exploit the optimization [11]. For example, when all the SIMD elements have a same value, the SIMD code can be reduced to scalar CPU code and run on regular CPU pipeline. In our experiments, as most taxi pickups happen in popular street intersections and they will be paired up with a limited number of road segments after spatial filtering (Section III.B), distance computation for the points in a grid cell are very likely to follow a same data path across the 16-way SIMD elements on the MIC accelerator which makes the optimization highly effective. As we shall see in Section IV.C (next subsection), the SIMD speedups are only about 5X for PIP due to different nature of data and computation.

The three GPUs have similar performance. GTX Titan (GPU3), which is based on the Kepler architecture, is about 13% better than GPU1 (Quadro 6000) and GPU2 (Tesla C2050), both are based on the Fermi architecture. GPU1 and GPU2 have very similar results as expected because the most significant difference between them is memory capacity (6GB and 3GB) which is not a limiting factor in the tests. For GPU3, although its number of processors is 6X higher (but about 30% slower with respect to clock frequency), it does not achieve the level of speedup as one would expect. This is primarily due to significant irregular data accesses to GPU global memory which make the experiments more memory bound. In contrast, for PIP experiments to be detailed in the next subsection (IV.C) where the experiments are more computing bound, the performance of GPU3 is 30% better than GPU1 and GPU2. We expect that Kepler-based GPUs will gain higher speedups over previous Fermi-based GPUs for more computing intensive applications.

When comparing the performance across the three CPUs, three GPUs and the MIC accelerator, from Table 2 we can see that, while GPUs are much faster over the previous generation CPUs (CPU1 and CPU2), the gap is much smaller for the new generation CPUs. In fact, CPU3 is roughly 2X better than the three GPUs in this particular application, when both multi-cores and VPUs are exploited. The MIC accelerator is even about 3X faster than GPU3. However, the GPUs are still 2X-5X faster if VPUs on multi-core CPUs and MICs are not utilized. Given that Intel Xeon Phi 3120A and GTX Titan are released around the same time with comparable price tags, it is fair to say that MICs are competitive in applications that involve significant irregular data accesses, provided that SIMD elements on VPUs are fully utilized with good optimizations from compilers like ISPC.

*C. Results of PIP Topoloigcal Test*

The PIP experiments on the 50 million species occurrence locations and 15 thousand complex polygons represent a different category of geospatial processing which is not only data intensive that involve significant irregular data accesses (as in P2P experiments reported in the previous sub-section) but also much more computing intensive. This is because, as discussed in Section IV.A, the average number of vertices in a polygon is about 280 and it is much higher than the average number of vertices in road segments in the NYC LION data set which is only a few in the worst case. Similar to what we have reported for P2P experiments, runtimes for the three CPUs, the MIC accelerator and the three GPUs are listed in Table 4. Likewise, the five speedups for the three CPUs and the MIC accelerators are listed in Table 5.

When comparing the results listed Table 4 and Table 5 with those listed in Table 2 and Table 3, respectively, we can see that the multi-core scalability is still close to linear for both CPUs and the MIC in PIP experiments which may indicate the advantages of our data parallel designs and its effectiveness in facilitating parallel libraries such as TBB for task scheduling. However, the speedups due to VPU accelerations are much lower than SIMD width. The speedups are only 5% - 42% on CPU1 and CPU2 with a SIMD width of 4, about 3X on CPU3 with a SIMD width of 8 and 5X on the MIC accelerator with a SIMD width of 16. While the speedups are still respectable, they are far below their respective SIMD widths, possibly due to two reasons. First of all, the large numbers of polygon vertices that points need to loop through may make it difficult to be cached fully on L1 cache. Given that the L1 cache size is only 32 KB per core for CPU1 and 256 KB per core for CPU2 and CPU3, and each polygon vertex takes 8 bytes (represented as two single precision floats), only 4K/16K vertices can be cached in L1 at maximum in CPU1 and CPU2, respectively. However, there are quite some polygons whose numbers of vertices are far above these numbers. Unfortunately, the computation that is related to these complex polygons typically dominates the overall computation as the number of operations per test is proportional to the ray-crossing algorithm we have used in PIP. Second, points (species occurrence locations) in this set of experiments are much more scattered (world-wide coverage) than in the previous set of experiments (taxi pickup locations in major street intersections), which may make the optimizations provided by the ISPC compiler less effective. As shown in the last row of Table 5, the combined multi-core and SIMD speedups are 8X-9X for CPU1 and CPU2, nearly 46X for CPU3 and more than 300X for the MIC accelerator. While the high MIC speedups are mostly due to the low performance of its weak cores (low clock frequency with in-order only execution) when only a single core is used (nearly 14X lower than CPU3), the 46X speedup that has been achieved by CPU3 can demonstrate the importance of good data parallel designs that scale with the number of processor cores and SIMD widths.

When comparing the runtimes across the seven types of processors listed in Table 4, it is clear that CPU3 is only moderately faster than CPU1 and slightly faster than CPU2 in PIP experiments in the single core without SIMD configuration. This is expected as the runtimes are proportional to their respective clock frequencies. It is interesting to observe that there are significant improvements of CPU2 over CPU1 (~2X) in PIP experiments while there are insignificant improvements of CPU2 over CPU1 in P2P using the single core without SIMD configuration. While more research is needed, a possible explanation is the availability of the 256 KB per-core L2 cache on CPU2 which is not available on CPU1. The per-core L2 cache may allow keep most frequently used polygon vertices in the cache which may significantly reduce memory traffic and improve the overall performance.

While the performance of GPU1 and GPU2 are still 2.5X to 5X better than CPU1 and CPU2 as we have reported previously [5, 15], CPU3 is now about 1.6X better than GPU3 when VPUs are utilized. This is somehow surprising, given that GPUs are frequently reported to have significant performance gains over CPUs [19], including our own comparisons using previous generations of GPUs and multi-core CPUs. One possible explanation is that newer generation multi-core CPUs are catching up and may help achieve higher performance for data-intensive applications with good data-parallel designs. On the other hand, we also expect that the newer generation GPUs may also significantly improve its performance for not only computing-intensive but also data-intensive applications. We note that the MIC accelerator in our experiments achieves very close performance when compared with GPU1 and GPU2 but lower performance than GPU3. Given that the Xeon Phi 3120A device we are using in experiments (P2P and PIP) is at the lower end of the product line, we expect that higher end MICs (e.g., 7120P [6]) may achieve comparable performance as GPU3, which is also the high end of Nvidia GTX GPU product line.

Table 4 List of Runtimes for Point-In-Polygon Test Experiments for all Seven Processors (in Milliseconds)

| Configuration | CPU1 | CPU2 | CPU3 | MIC | GPU1 | GPU2 | GPU3 |
|---|---|---|---|---|---|---|---|
| SC | 459268.06 | 242363.52 | 226044.27 | 3260769.25 | | | |
| SC+SIMD | 389033.25 | 169035.75 | 63549.51 | 616230.50 | | | |
| MC | 57787.19 | 31756.97 | 16313.82 | 54070.88 | | | |
| MC+SIMD | 55233.37 | 25979.75 | 4925.65 | 10805.37 | 10857.50 | 10866.29 | 7831.93 |

Table 5 List of Speedups for Point-In-Polygon Test Experiments for Three CPUs and MIC

| | Configuration | CPU1 | CPU2 | CPU3 | MIC |
|---|---|---|---|---|---|
| Multi-core Speedup | SC/MC | 7.95 | 7.63 | 13.86 | 60.31 |
| | (SC+SIMD)/(MC+SIMD) | 7.04 | 6.51 | 12.90 | 57.03 |
| SIMD Speedup | SC/(SC+SIMD) | 1.18 | 1.43 | 3.56 | 5.29 |
| | MC/(MC+SIMD) | 1.05 | 1.22 | 3.31 | 5.00 |
| Overall Speedup | SC/(MC+SIMD) | 8.32 | 9.33 | 45.89 | 301.77 |

*D. Further Discussions*

While quite some comparisons have been presented in the previous two sub-sections, we would like to dedicate this sub-section for additional cross-application comparisons and further discussions. When comparing the results listed in Table 2 and Table 4, we can see that the best performance is achieved by the MIC accelerator (Xeon Phi 3120A) for P2P experiments and by CPU3 (dual Xeon E5-2650V2) for PIP experiments, when both multi-core and VPUs are combined. The low floating point intensity and simpler control logic in P2P experiments may make it more suitable for MICs with weaker cores but larger number of cores and higher memory bandwidth.

While the best performance (among both sets of experiments) is 2X-3X better than those of GPUs, it is 1.5X-5X worse than that of GPUs without using VPUs, even on the most recent generation CPUs. Our experiments clearly demonstrate the importance of utilizing SIMD computing power on VPUs that come with CPUs. The better performance of CPU3 and the MIC accelerator over GPUs also indicate the importance of large caches for data-intensive computing with significant irregular data accesses. The experiment results may be useful for developing future heterogeneous computing architectures.

It is increasingly popular to exploit multi-core computing power using various parallelization tools (e.g., TBB [10]). Porting serial code to multi-cores in a straightforward manner by running multiple independent tasks on multiple cores requires only moderate efforts. However, utilizing SIMD computing power on VPUs is still technically challenging which may require significant re-designs and re-implementations. From a application developer perspective, while learning GPU programming using CUDA or OpenCL already has a steep learning curve, currently programming VPUs requires manipulate SIMD intrinsic functions directly, which has a much deeper learning curve for the reasons discussed in Section II. CUDA-based programming model allows each thread have its own registers and local variables and control logics such as branching are very similar to traditional scalar programming. However, programming SIMD intrinsic functions, where branching requires use mask registers (as all elements share same registers), is much more complex and error-prone. Furthermore, forcing developers to call different SIMD intrinsic functions for different data types make it unproductive and unattractive to developers. While major compilers allows to use pragma based directives to facilitate automatic simdificaiton (or auto-vertorization), except for simple loops over array elements, the achievable performance gains typically are not significant, due to the difficulties in expressing complex semantics in pragma directives. We thus prefer explicitly express parallelisms in programs in a way similar to what we have done for CUDA and ISPC implementations of the two selected spatial operators.

While there are a few pioneering work on exploiting SIMD computing power for relational data management with limited scope (e.g. [20, 21, 22]), to our knowledge, we are not aware of previous works on exploiting SIMD computing power on VPUs for geospatial processing. Encouraged by the good performance of the ISPC-based SIMD programming on VPUs for P2P that has demonstrated comparable performance on previous generation of CPUs [4], our experiments on ISPC-based SIMD programming on VPUs for PIP based geospatial processing also have achieved similar good results. The performance on the current generations of CPUs and MICs is even better than that on the current generation of GPUs. As such, although we are aware that the complier (ISPC in specific) driven approach may not achieve the highest possible SIMD computing power on VPUs when compared with programming intrinsic functions directly, we advocate to take advantage of DSL compliers (such as ISPC) that allows program application logics in a scalar way and translates the programs into SIMD code using either source-to-source or source-to-binary approaches. From a practical perspective, the similarities between CUDA and ISPC code may further make them compatible in the future. We are in the process of developing more data-parallel designs for spatial operations and implement them using both CUDA on GPUs and ISPC on CPUs. Additional results will be reported in our future works.

## V. Conclusions and future work

In this study, we have experimented the designs and implementations of two popular spatial operators, namely point-to-polyline shortest distance computation and point-in-polygon topological test, on three CPUs, a MIC accelerator and three GPUs using real large-scale geospatial data. Our experiments results have shown that, while GPUs are significantly faster than multi-core CPUs without utilizing VPU SIMD computing power, the performance of the current generation multi-core CPUs with combined VPU processing power can be several times better than GPUs. Our data parallel designs, Intel TBB based implementations for multi-core CPUs and MICs, ISPC based implementations for VPUs, CUDA based implementations for GPUs, and extensive experiments can serve as geospatial domain case studies for performance evaluations and contribute to defining future parallel computing hardware architecture, language tools and runtime libraries.

For future work, as discussed inline, first of all, we would like to investigate further on the significant performance gains of newer generation CPUs over the previous ones in P2P experiments by taking more hardware, system and data related factors into considerations. Second, we plan to further optimize MIC performance by considering more hardware-specific features and investigate scalability of TBB-based scheduling for much larger number of threads. Finally, we would like to design and implement more spatial operations (e.g., K- Nearest Neighbor and shortest paths) on multi-core CPUs, VPUs and GPUs as well as their hybridizations for more comprehensive evaluations and comparisons.